\documentclass[12pt]{article}
         
         \usepackage{latexsym}
         \usepackage{amssymb}
         \usepackage{graphicx}

           \newtheorem{theorem}{Theorem}[section]
           
           \newtheorem{definition}{Definition}[section]

\begin{document}

\begin{center}
 {\Large{\bf
Tectonic plate  under  a localized boundary \\stress:
fitting of a zero-range solvable model.}\\
 \vskip0.3cm
L. Petrova$^1$,  B. Pavlov $^{1,2}$}.
\end{center}
\vskip0.3cm
 $^1$ V.A.Fock  Institute of Physics, St. Petersburg
University, Russia.

\noindent
 $^2$ Department of Mathematics,
the  University of Auckland, New Zealand. \vskip0.3cm

\begin{center}
{\bf Abstract}
\end{center}
We suggest a method  of  fitting   of a zero-range  model of a
tectonic plate under a boundary stress on the basis of comparison of
the theoretical formulae for the corresponding
eigenfunctions/eigenvalues with the results  extraction under
monitoring, in the remote zone,  of  non-random (regular)
oscillations of the Earth with periods  0.2-6 hours, on the
background seismic process, in case of low seismic activity.
Observations  of changes of the characteristics of the oscillations
(frequency, amplitude and polarization) in course of time, together
with the theoretical analysis of the fitted model, would enable us
to localize the stressed zone on the boundary of the plate and
estimate the risk of a powerful earthquake at the zone.

 Key-words {\it Tectonic plate, Zero-range interaction, Operator
extension}

PACS numbers  91.45.D, 02.30.Jr, 02.30.Tb

\vskip0.3cm
\section{ Dynamics of the  system of  tectonic plates\\
and the motivation of  the  zero-range model.}

The lithosphere of Earth consists of 14 tectonic plates which jigsaw
fit each other. The plates are isolated from underlying  solid
structures within the Earth mantle by the low-viscosity layer of the
asthenosphere  which is  formed, due to various conditions,
including high pressure and temperature, in the interval of depth
100 -200 km. The plates move on the surface of Earth, due to
convective flows and variations of the angular speed  of Earth,
gliding  on the low-viscosity layer of asthenosphere and interacting
with each other at some {\it active boundary zones}, see  below.
\begin{figure}
[ht]
\begin{center}
\includegraphics[width= 6in]{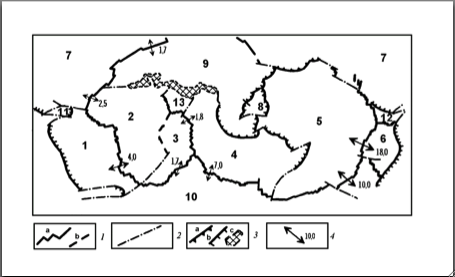}
\end{center}

\caption\,\,\, Boundaries of tectonic plates: 1- divergent
boundaries (a - oceanic ridges, b- continental rifts), 2-
transforming  boundaries, 3-convergent  boundaries (a-insular, b-
active continental outskirts, c - collisions of plates). Directions
and  velocities  of  movement  of plates  ( cm/ year).
\label{F1:figure}
\end{figure}
Fig. 1 below  {\footnote { Fig.1 is  borrowed  from the book  of  S.
Aplonov, \cite{Aplonov2001}, and is included in our text with
permission of the Publishing House of the St.  Petersburg
University.}} shows a complex form of boundaries between tectonic
plates  and their movements in different directions. The  plates are
enumerated  in the  following order :  1. South-American plate,\,\,
2. African plate ,\,\,
 3. Somali plate,\,\,
 4. Indian and  Australian  plates,\,\,
 5. Pacific plate,\,\,
 6. Nazca plate,\,\,
 7. North-American plate,\,\,
 8. Philippines plate,\,\,
 9. Euro-Asian plate,\,\,
 10. Antarctic plate,\,\,
 11. Caribbean plate,\,\,
 12. Cocos plate,\,\,
 13. Arabian plate.\,\,

In addition to the 14 large plates  described  above  there are
 38 smaller  plates (Okhotsk, Amur, Yangtze, Okinawa,
Sunda, Burma, Molucca Sea, Banda Sea, Timor, Birds Head, Maoke,
Caroline, Mariana, North Bismarck, Manus, South Bismarck, Solomon
Sea, Woodlark, New Hebrides, Conway Reef, Balmoral Reef, Futuna,
Niuafo'ou, Tonga, Kermadec, Rivera, Galapagos, Easter, Juan
Fernandez, Panama, North Andes, Altiplano, Shetland, Scotia,
Sandwich, Aegean Sea, Anatolia, Somalia), for a total of 52 plates.

It is  commonly  accepted that  the tectonic plates  are relatively
thin elastic structures, approximately 100 km. thick, with linear
size from 1000 km to several thousand  km. The  material of the
plates, at the depth 100 km., has typical Young's modulus
17.28\,$\times$\,10$^{10}$ {\it kg\,\,m}$^{-1}$\,{\it sec}$^{-2}$ ,
density 3380 {\it kg\,m}$^{-3}$  and Poisson coefficient 0.28. The
velocity of the longitudinal waves in these materials is
approximately  8000 {\it m\, sec}$^{-1}$, and the velocity of the
transversal (flexural) eigen-waves depends on  the eigenvalues and
varies, depending  on the type of the wave, on a wide range around
4500 {\it m\, sec}$^{-1}$, see the formula (\ref{flexural_speed})
below. Because of non accurate matching of the boundaries of the
neighboring plates, the zones of direct contact of the plates are
typically small- about 100 km.- compared with the linear size of the
plates. Remaining inter-plate space is filled with loose materials,
which are not able to accumulate any essential amount of elastic
energy caused by the deformation. These materials can damp
oscillations with short periods  as  20 min. - 1 hour. Damping of
acoustic waves by  loose materials was  discussed in
\cite{Nigmatulin,Kuperin,XL01}. Since tectonic plates are relatively
thin, a major part of their kinetic energy is stored  in the form of
oscillatory flexural modes. The underlying layer of the
asthenosphere makes flexural oscillation of plate possible, but also
helps damping of the flexural waves with short periods, due to
non-zero viscosity. Based on above data we conjecture that the
flexural modes in different tectonic plates, in a certain range of
periods, supposedly between  $0.2 h - 1 h$, see below  the
estimation of  periods  of eigen-modes of a model rectangular plate,
are elastically disconnected from each other. Thus we expect that
these flexural modes characterize elastic properties of the plates,
but not the global elastic properties  of the Earth's crust.

The  convective  flows in the asthenosphere  and  variations of the
angular speed of Earth, thanks to long-time variations of the moment
of inertia of Earth,  may cause  collisions of neighboring plates.
In presence of the liquid friction on the underlying layer of the
asthenosphere, variations of the angular speed of earth cause mutual
displacements of the neighboring plates, because small plates react
immediately on the variations of the angular speed, and larger
plates lag behind. These displacements cause collisions of the
plates in the active zones, where the plates directly contact each
other. Generally, when the angular speed of Earth decreases, with
growing of the moment of inertia  due to displacement of the center
of gravity of Earth,  the collisions may  occur on the eastern
boundaries of the major plates contacting smaller plates, observe,
for instance, the contact of the Euro-Asia plate and  the
Philippines plate on Fig. 1. The stress caused by the collision may
be either discharged due to forming cracks in the plates, splitting
the active zone into independently moving fragments, or, being
applied for an extended period of time, may cause accumulation of a
considerable amount of (potential) elastic energy in the active
zones of contacts, in form of elastic deformation of the stressed
plates. This  energy may be eventually discharged  in form of a
powerful earthquake.

 Accumulation of the elastic energy, due to standard variational
principle \cite{CH2}, causes the {\it increment of eigenfrequencies
} of the tectonic plates. In \cite{PLZ88,P2000,P2002} oscillatory
processes, with similar spectral characteristics, were  observed  in
mutually remote zones of the Euro-Asia  plate. The  authors of
\cite{PLZ88} suggested  calling the processes {\it
seismo-gravitational oscillations of the Earth} ( SGO). Recent
analysis, based on data of the international GEOSCOPE network,
revealed the  existence of {\it global} oscillations with periods
3.97 h, 3.42 h, 1.03 h and 0.98 h. SGO with smaller periods do  not
have global character. They are usually observed in certain plates.
For measurements of SGO, special devices are used (``vertical
pendulum''), with high sensitivity to the variations of amplitudes
and frequencies of SGO within the interval of periods 0.2 - 2 hours.
Modification of the registering channel of the device allowed us to
extend the interval to 0.2-6 hours. Results of  analysis of
measurements of SGO with relatively short periods from this range
are presented in \cite{Petrova_FZ08}. The data on  SGO with  longer
periods and characteristics of modern 3-channel seismographs can be
found in \cite{{Petrova_FZ07}}. Observations with these devices
reveal a wide spectrum of SGO. Some of the  measured frequencies of
SGO coincide with short-time variations of the angular speed of
Earth, which were estimated based on astronomical observations, see
the \cite{Petrova_FZ06}. All experimental data confirm the presence
of active energy in the system of tectonic plates, relevant to the
inhomogeneity of the lithosphere.
\begin{figure} [ht]
\begin{center}
\includegraphics[width= 4in]{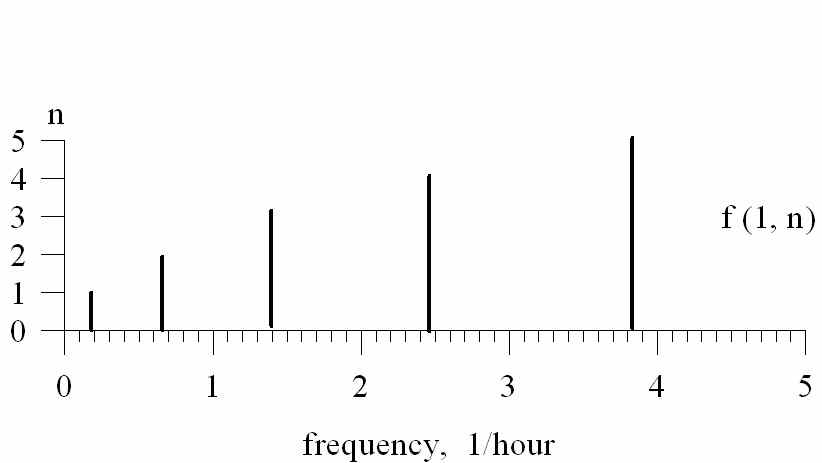}
\end{center}
\caption\,{ Computed eigenfrequencies of the rectangular plate 4 000
km $\times$ 8 000 km for the modes f(1, N)} \label{F2:figure}
\end{figure}
Direct calculations of eigenfrequencies of the flexural modes of a
rectangular thin plate  $\Omega = d_1 \times d_2 $, $d_1 =$ 4 000
km, $d_2 =$ 8 000 km, 200  km thick,  were done, see \cite{Thesis},
based on bi-harmonic model, under natural assumptions concerning the
density of the material is 3380 kg/m$^3$, the Young's modulus is
17.28 $\times$ 10$^{10}$ $ kg\,m^{-1}\,sec^{-2} $, and the Poisson
coefficient is equal to 0.28. Though the Young's  modulus of a
tectonic plate varies  on a  wide range of values, it is possible to
provide a  rough estimation, see below, of the periods of  SGO and
the length of the corresponding running waves of  a tectonic plate,
by computing these parameters for a  model rectangular plate with
elastic  parameters  equal  to those  of the tectonic plate on the
half depth, 100 km.  Indeed, the corresponding dynamical equation
for the transversal (vertical) displacement $f$ is
\begin{equation}
\label{dynamic} \rho \frac{\partial^2 f}{\partial t^2} + \frac{h^2\,
E}{12 (1 -\sigma^2)}\Delta^2 f = 0.
\end{equation}
In  \cite{Thesis} the  simplest Neumann boundary conditions
$\frac{\partial u}{\partial n} \bigg|_{\partial \Omega} =
\frac{\partial \Delta u}{\partial n} \bigg|_{\partial \Omega} = 0 $
are imposed  on the boundary. This  allows us  to  solve the
dynamical equation by Fourier method, via  separation of  variables.
The eigenfunctions of the bi-harmonic operator coincide with  the
eigenfunction of the Neumann Laplacian, $ f_{l_1,l_2}= \cos
\frac{\pi l_1 x_1}{d_1} \cos \frac{\pi l_2 x_1}{d_2}$, but the
eigenvalues of the  Neumann bi-harmonic  operator
\[
\lambda^{\Delta^2}_{l_1,l_2} =
\left[\lambda^{\Delta}_{l_1,l_2}\right]^2 = \left[\pi^2
\left(l^2_1\, d_1^{-2} + l^2_2\,\, d_2^{-2}\right) \right]^2
\]
are squares  of  the corresponding eigenvalues of the Neumann
Laplacian. Taking into account  that
\[
\frac{h^2\, E}{12 \rho(1 -\sigma^2)} = 0.167 \,\, 10^{20} = \left[
0.64 \,\times 10^5\right]^4 =: \alpha^4,
\]
we  find the periods of flexural eigen-ocsillations  from the
formula:
\[
\omega^2_{l_1,l_2}= \left( \frac{2\pi}{T_{l_1,l_2}}\right)^2 =
\alpha^4 \,\left[\lambda^{\Delta}_{l_1,l_2}\right]^2,
\]
or
\begin{equation}
\label{momentum} \omega_{l_1,l_2}= \frac{2\pi}{T_{l_1,l_2}} =
\alpha^2 \,\,\,\lambda^{\Delta}_{l_1,l_2}=:\alpha^2\,\,
k^2_{l_1,l_2},
\end{equation}
where $k$ plays the role of  the corresponding  momentum. The  speed
of the transversal ( flexural) waves is calculated as
\begin{equation}
\label{flexural_speed}
 v^f_{l_1,l_2} = \frac{\partial \Omega}{\partial k} = 2\,\,
 \alpha\,\,
k_{l_1,l_2}.
\end{equation}
Then for the above model  plate the periods of flexural oscillations
are defined by the formula $T_{hours} = 2.63\,\, \,[4 l_1^2 +
l_2^2]^{-1}\,\,\, hours $, and the  velocity of the corresponding
flexural waves are calculated as $v^f_{l_1,l_2} = 3400 \sqrt{4 l_1^2
+ l_2^2}\,\,\, m \,\,sec^{-1}$. In particular, the period of the
flexural oscillation $f_{1,2}$  and the speed of the  corresponding
flexural wave $v^f_{1,2}$ are calculated as
\begin{equation}
\label{flexural_speed} T_{1,2} = 0.33\,\, hours,\,\,\, v^f_{1,2} =
9520\,\, m \,\,\sec^{-1},
\end{equation}
and  less  for  longer periods. The  corresponding space-temporal
oscillatory mode is
\[
\cos{\omega_{1,2} t} \cos\frac{\pi \, x_1}{4000}\,\cos\frac{\pi \, 2
x_2}{8000},
\]
with $x_1,x_2$ measured in kilometers.  This oscillatory mode can be
represented as a linear combination of running  flexural waves
\begin{equation}
\label{running} \cos\left[\omega_{12}t \pm \frac{\pi \, x_1}{4000}
\pm \frac{\pi \, x_2}{4000}\right].
\end{equation}
The corresponding  wavelength is  estimated by the minimum of
$\sqrt{\Lambda_1^2 + \Lambda_2^2}$, where $\Lambda_1,\Lambda_2$
correspond  to the  shift of the  running  wave  by  the
corresponding  period in time. For  instance
\[
\omega_{1,2} T_{12} = 2\pi  = \frac{\pi \, \Lambda_{1} + \pi
\Lambda_{2}}{4000}
\]
gives an estimation of the  minimal wavelength  as
$\sqrt{\Lambda_1^2 +\Lambda_1^2} \geq 5656 $ km,  which is  much
more than the diameter of the active zone, 5656 km  $>>$  100 km.
Hence the  zero-range model  can be  used, under above  assumption,
for  the model  stressed plate.

It appeared that periods of the  eigen-modes $f_{1,1} - f_{5,5}$ sit
in the interval 0.21 h - 5 h and their total number and distribution
looks similar, see Fig. \ref{F2:figure}, to SGO described in the
paper \cite{PLZ88,P2000,P2002}, despite the trivial plane
rectangular geometry of the plate and trivial Neumann boundary
conditions.

Examination of  the results observed  in \cite{PLZ88,P2000,P2002}
 and theoretically obtained in \cite{Thesis} resulted in the
conjecture that transversal  SGO {\it can be interpreted as flexural
eigen-modes} of the relatively thin tectonic plates.
 We  hope that  the  zero-range  model may be also
used in this  periods range  for real  tectonic plates as well.

In the next section we model dynamics of tectonic plates based on
the bi-harmonic boundary problem with ``natural'' boundary
conditions.
\begin{figure}
[ht]
\begin{center}
\includegraphics[width= 4in]{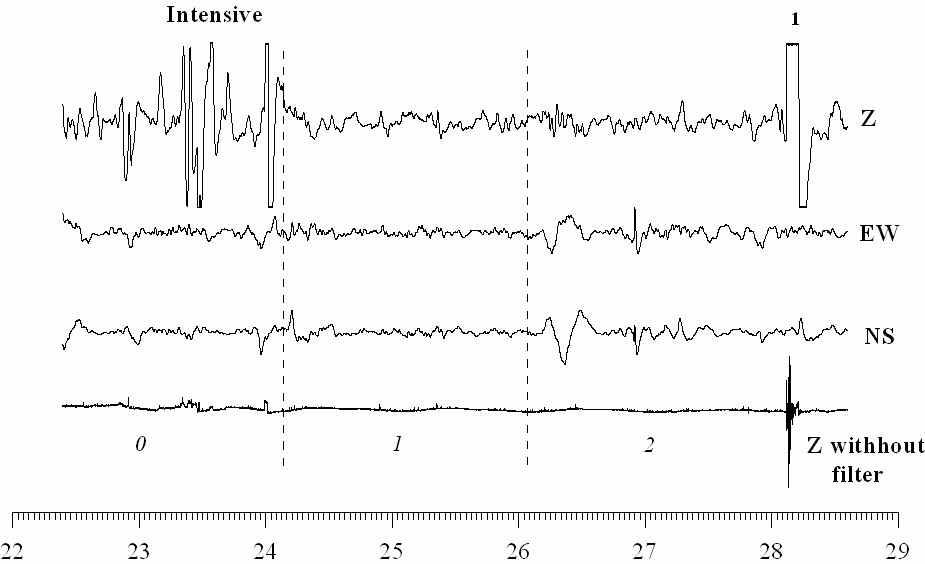}
\end{center}
\label{F3:figure} \caption {Dynamics of seismo-gravitational
oscillations observed in St. Petersburg, 22-29 March 2000, see the
text below.}
\end{figure}

Variation of the  character of SGO is  seen from comparison of  the
graphs of the amplitudes of SGO  on  48-hours intervals of time
separated by the vertical dotted lines.  The  graphs on these
intervals show the reaction of the device on the  oscillations of
the  ground.

\begin{figure}
[ht]
\begin{center}
\includegraphics[width= 3in]{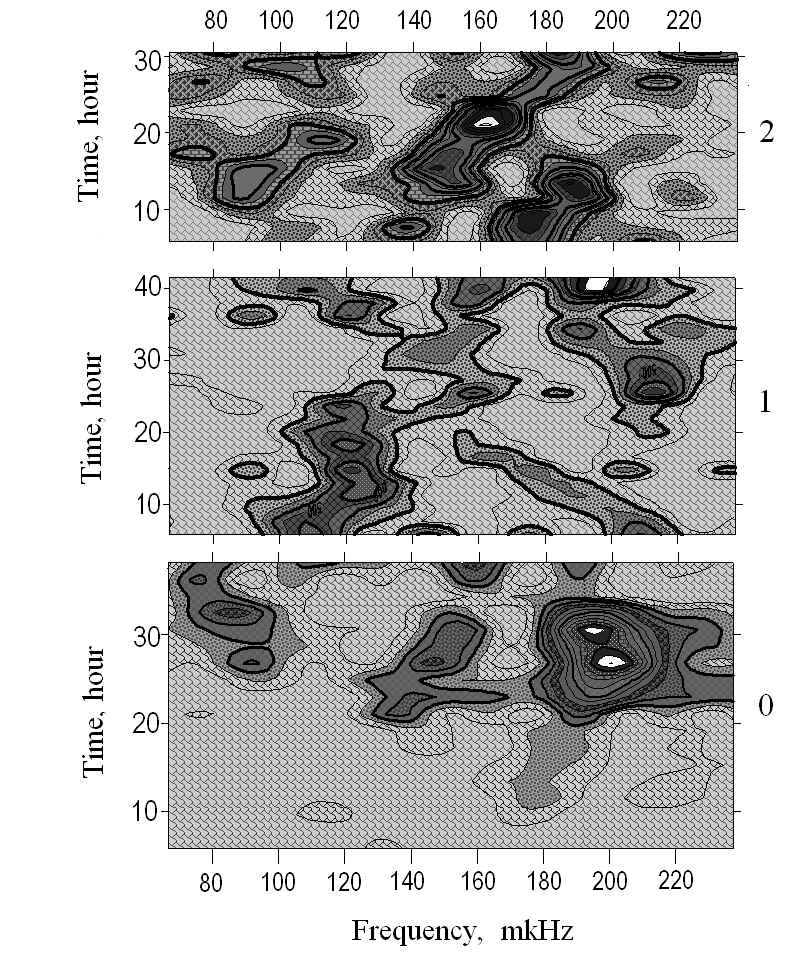}
\end{center}

\caption{ Time-frequency analysis of the vertical component of  SGO
} \label{F4:figure}
\end{figure}

A typical phenomenon of intense seismo-gravitational {\it pulsation}
(SGP) was registered on the vertical component $Z$ on the initial
interval marked by 0. This phenomenon was also noticed in the
earlier paper \cite{LPO90}. See more about  SPG in  the further
text, after the  figure \ref{F4:figure}.

The components Z,EW and NS are obtained after  filtration. The graph
on the interval  1 shows the reaction of the filter  on the maximal
phase of the earthquake 28 March 2000 in Japan. The non-filtered
graph Z represents the reaction of the base of the vertical
seismograph on the transversal oscillations of the ground.

Spectral-time cards, see below fig. \ref{F4:figure} represent the
effect of growing of frequencies of SGO  in  visual form. The data
of the observations were filtered by the band filter with the band
strip [60 min.-300 min.]. We used the gliding time window length 700
min., and step-wise shifts with the 5 minutes steps. The interval of
frequencies, measured in micro-Hertz,  was chosen as $[ 70 mcHz,250
 mcHz] =:  [F_{min}, F_{max}]$. For resolution $0.034 mcHz$ the
steps were halved to  guarantee  better smoothness of the spectral
function.

Effect  of  growing of the frequency of  SGO was detected only  on
spectral-time cards (ST-cards) of the  vertical component of SGO,
see  below  fig. \ref{F4:figure}. The frequencies are marked on the
horizontal axes,  time in hours - on the vertical axes, for the
intervals 0,1,2 respectively. The level of spectral amplitudes is
represented by  the variation of the color. Oscillations with large
spectral amplitudes, higher than the average level on the spectrum,
are marked by grey color. The maximal amplitudes are marked by white
color. The  values of these amplitudes on the interval 0 are 3.5
times  greater that  the average amplitude, ( frequency about 200
mcHz). On the interval 1 they exceed the average amplitude in 1.8
times (the  frequency about 200 mcHz). On the interval  2 they
exceed the average amplitude in 3 times ( the frequency about 160
mcHz). For SGO with frequencies 90-110 mcHz and 170- 190  mcHz the
amplitudes are 1.75 and 2.75 is times  larger than the average
amplitude. All three parts 0,1,2 of the fig. \ref{F4:figure} reveal
two patterns  of  inclined  gray domains  ( left-down to  right-up
and left-up  to  right-down) which correspond to SGO with growing
and decreasing frequencies, correspondingly. This patterns provide
an evidence of increment of the stored elastic energy in the  system
and discharging the  stored elastic energy, respectively in form of
oscillation modes with certain frequencies.

Growing of the  frequency on the interval 2:

1. The  frequency is  growing from  $f_{min} =  80 mcHz$ to $123
mcHz$ during the period of 15 hours.

2. The  frequency is  growing from  $f_{min} =  130 mcHz$ to $185
mcHz$ during the period of 28.5 hours.

3.The  frequency is  growing from  $f_{min} = 170 mcHz$ to $ 200
mcHz$ during the period of 8 hours.

Growing of the  frequency on the interval 0

1. The  frequency is  growing from  $f_{min} =  165 mcHz$ to $205
mcHz$ during the period of 33 hours.

2. The  frequency is  growing from  $f_{min} =  125 mcHz$ to $165
mcHz$ during the period of 15 hours.

Growing and  decreasing of frequencies of SGO are  easily noticeable
also on the interval 1.

Growing of the  frequency  is  characterized by  the ratio
$\frac{\Delta \nu}{\tau}$, where $\Delta \nu$ is the increment of
 the frequency and $\tau$ is the corresponding time interval, when
 the growing was  observed.

Other domains, where the frequency of the modes decrease with time
growing,  may be  interpreted  as  an evidence of local relaxation
of the stress, probably caused by local  destruction of the plate
(forming cracks).

We conjecture that the extended growth of frequencies of the SGO may
be considered as a precursor of strong earthquakes. Our ability to
extract  useful information from the observations of frequencies and
the shape of SGO is limited by our understanding of the mechanism of
the variation of the frequency and the shape  of SGO modes arising
from the boundary stress on the tectonic plates.

There were also  other observations of the shift  of frequencies of
selected modes of SGO. Note, that in numerous observations on SGO in
Leningrad (St. Petersburg) intense short pulses were also recorded.
They are  constituted by  several sinusoidal harmonics  with periods
from 30 minutes to 1 hour, and  the  total duration of the process
6- 10 hours. In several cases they were also followed by powerful
earthquakes - in 2-4 days.  This process was noticed first in
\cite{LPO90} and was given the name of ``seismo-gravitational
pulsations'', SGP. Statistical analysis was done based on the  data
of the 6 months monitoring of SGO in St-Petersburg. This  analysis
confirmed that the connection between SGP and the subsequent strong
earthquake is not random, with probability 95 \%.

Essential information on SGO  is  obtained  from  the  observation
of variations of the  frequencies and  the shape of  the
corresponding modes  in the  remote  zone. The typical  size of the
zone of contact is negligible - ``point-wise''- when compared with
the wavelength, which generally can be
 estimated  as  $\left[{\lambda}^{\Delta}\right]^{-1/2}$,
 see (\ref{running}).

Although  the boundaries of  real tectonic plate are not smooth,
however details of the local geometry of the boundaries may be
neglected compared with the wavelength  of typical  waves on the
plates. Then for  SGO with periods  20 min. - 1  hour we  may base
the theoretical analysis of SGO  on the bi-harmonic model for the
relatively thin plate with natural boundary conditions. In this
paper we also neglect  presence of the liquid layer of the
asthenosphere underlying the plate, thus reducing the problem to
construction  of self-adjoint  extensions of the 2-d bi-harmonic
operator.

In  \cite{PPI06} we  use the fact  that characteristics  of the
seismo-gravitational oscillations  in remote  zone depend on  a {\it
small  number  of basic parameters}.  On one hand, his  fact is can
be interpreted  in a spirit of  the Saint-Venant principle
\footnote{We are  grateful to Doctor Colin Fox for inspiring
discussion  concerning  the  Saint-Venant principle}. On the other
hand we  were able  to  interpret this fact  in the  spirit of
operator extensions.  Based on this observation we developed in
\cite{PPI06} a preliminary version of the above arguments  and
suggested  to use a solvable zero-range model of the tectonic plates
under a point-wise boundary stress, caused  by  the collision of
plates. The role of the parameter of the model was played by some
 $3\times3$ real Hermitian matrix  $M$, see also next section.
Thus the number of  the {\it Saint-Venant parameters} for the
boundary stressed relatively  thin  tectonic plate is 6. We presume,
that the matrix $M$ defines the type of the stress, depending on
mutual positions of the contacting plates at the active zone. There
is a good reason to  call  $M$ the Saint-Venant matrix and  the
number of the Saint-Venant parameters (6) - the Saint-Venant number.
We assume that mutual positions of the plates in active zone, and
hence the matrix $M$ remains essentially unchanged during extended
period. Thus the matrix $M$ characterizes the type of  contact of
tectonic plates  in this  location and remains  the same for all
earthquakes arising from the given active zone. In this  paper,
based on  the {\it compensation of singularities} in the fundamental
Krein formula, we suggest an explicit formula for the perturbed
eigenfunctions of the plate. Comparison of the calculated
eigenfunction  with the data of the instrumental observations in
remote zone, permits, in principle, to {\it fit} the model, that is,
to find the basic matrix $M$. Once {\it fitted}, the constructed
model would allow to calculate in explicit form the increment of the
eigenvalues and the variations of the shape of the eigenfunctions of
the stressed (perturbed) plate depending on the type and the
magnitude of the local  stress.  We assume that  the shifts of the
eigenfrequencies and the changes of the  shape of the eigen-modes
can be  measured for each active zone. With numerous active stations
in the  GEOSCOPE  and IRIS  networks, fitting of the proposed model
can be done, eventually, for all active zones on the boundary of
each major tectonic plate.

Generically earthquake  hits only one active zone at a time. Then
comparing the observed  shift of the eigenfrequencies and the
variations  of the  shape  of the mode in the  remote  zone, where
the Saint-Venant principle  is applicable, with the  results of
computing  based on the  model, we will be able to localize the
excited  active zone   based  on the observed changes  of the
eigen-frequencies and the shapes (amplitude, polarization) of the
flexural eigen-modes.

Mathematically  the  zero-range  model of the isolated  point-wise
stressed tectonic plate  and  a similar  zero-range  model of the
point-wise stressed  tectonic plate  submerged into  environment
formed by other  plates, intermediate  layers and  asthenosphere,
differ  by the  type  of basic  equations, but have a  lot in
common. In particular, the number of  free parameters (6) for the
point-wise active zone, in bi-harmonic model and  Lame  model  is
the same. We interpreted these parameters in the spirit of the
Saint-Venant principle as essential parameters describing the shape
of the wave-process in the remote zone, see \cite{PPI06}. In this
paper we obtain  the first  order approximation for the perturbed
eigenfrequencies and eigenfunctions based on  a {\it modified} Krein
formula, see \cite{MP2007,APY_INI_07}, for  the point-wise stressed
thin plate described by the bi-harmonic equation. The explicit
representation of the perturbed eigen-modes permits to {\it fit} the
zero-range model suggested in \cite{PPI06} based on results of
instrumental measurements of SGO. We also conjecture that the
properly fitted solvable model of the stressed tectonic plate may
help to enlighten the nature of  the pulsations. \vskip0.3cm

\section{Zero-range model of the  point-wise boundary stress}

We may  base our  approach on  the standard  mathematical model of
the tectonic plate in form of a thin  elastic  plate, thickness $h$,
with free edge, or on  the  3-d  Lame  equations  for displacements.
We  consider  both options, describing in the next two subsections
specific details of both models. Then we  develop  the  common part
of the theory, for both models simultaneously.

\subsection{ Thin plate model for the isolated tectonic \\plate under
the  localized boundary stress}

Denoting by $D$ the ``the  bending stiffness ", connected with Young
modulus $E$, the thickness $h$ of the plate and the Poisson
coefficient $\sigma$ by the formula $ D =Eh^2 [12
(1-\sigma^2)]^{-1}$, we represent, following
\cite{Landau_Elasticity}, the corresponding dynamical equation for
the normal displacement $u$ as
\[
\rho  \frac{\partial^2 u }{\partial s^2}= D \Delta^2 u.
\]
We  consider  time-periodic solutions  $u (\omega s,\, x)$of the
 equation and  separate  the time, thereby reducing the  dynamical problem
 to the  spectral problem  with the spectral parameter
\begin{equation}
\lambda = \rho  D^{-1}\,\, \omega^2
\end{equation}
for  a bi-harmonic operator on a compact 2-d domain $\Omega$ - the
tectonic plate - with a smooth boundary $\partial \Omega$:
\[
A u = \Delta^2 u = \lambda u,\, u \in  W_2^4 (\Omega),
\]
and  free boundary condition involving  the tangential and normal
derivatives of  the  displacement $u$ and the tension  $\Delta u$:
\begin{equation}
\label{free_bound_cond}  \left[\frac{\partial \Delta u}{\partial n}
+ (1- \sigma)\frac{
\partial^3 u}{\partial n \partial t^2 }\right]
\bigg|_{\partial \Omega}= 0,
\]
\[
\left[ \Delta u - (1-\sigma)\frac{\partial^2 u}{\partial t^2}\right]
\bigg|_{\partial \Omega}= 0.
\end{equation}
Here   $n,t$ are the normal and the  tangent directions  on the
boundary.

The bi-harmonic operator $A$ is selfadjoint in the Hilbert space
$L_2 (\Omega):= \mathcal{H}$. The  eigenfunctions of $A$ are  smooth
and they form an orthogonal basis in $L_2(\Omega) ={\mathcal{H}}$.
We consider the restriction $A_0$  of $A$ onto $D(A_0)$ constituted
by all smooth functions vanishing near the boundary point $a\in
\partial \Omega$. The restriction is symmetric, but  it is not
selfadjoint, because the range of it $\left( A - {\lambda} I
\right)D(A_0) $, for complex $\Lambda$ has a nontrivial complement
$N_{\lambda}$  which is a linear hull of the  Green function $G
(x,a,\bar{\lambda}):= g_0 (x,\bar{\lambda})$ and its tangential
derivatives $\frac{\partial G (x,a,\bar{\lambda})}{\partial t}:= g_1
(x,\bar{\lambda}),\, \frac{\partial^2 G
(x,a,\bar{\lambda})}{\partial t^2}:= g_2 (x,\bar{\lambda})$ of the
first and second  order, at the point $a$. The orthogonal complement
$N_{\lambda}$ of the range is  called the
 ``deficiency subspace'', and  elements of it  - ``deficiency
elements'':
\[
N_{\lambda}: = {\mathcal{H}} \ominus (A_0 - \lambda I) D_0 =
\bigvee_{s=0}^2 g_s (*,\bar{\lambda}).
\]
The  deficiency elements  have, at the boundary point $a$,
singularities of different types ( see  for instance \cite{Mazja77},
where  much  more  general problem is  considered) :
\[
g_0 (n,t)\approx (n^2 + t^2) \ln (n^2 + t^2),\,\, g_1 (n,t) \approx
t\ln (n^2 + t^2),\,\, g_2 (n,t) \approx \ln (n^2 + t^2),
\]
hence they are linearly independent and  form a  basis in the
deficiency  subspace. The deficiency subspace at the spectral point
$\bar{\lambda}$  is
\[
N_{\bar{\lambda}}: = {\mathcal{H}} \ominus (A_0 - \bar{\lambda}I)
D_0 = \bigvee_{s=0}^2 g_s {*,{\lambda}}.
\]
The  dimensions  of the  deficiency subspaces $(3,3)$ constitute the
``deficiency index''. Hereafter we  select $\lambda = i$ and attempt
to construct  a  self-adjoint  extension  of  $A_0$, which will play
a role of a  zero-range model of the  tectonic plate under the
boundary strain.

Note  that  Lame  model the  deficiency index is also  $(3,3)$, on a
smooth boundary. The  role  of  deficiency elements  is  played  by
the columns  of the Green matrix. The boundary  of  the  tectonic
plates may be  assumed  smooth for the
 long  waves (small $\lambda$), since  the integral  shape of  the solutions
 of  the differential equations with small $\lambda$ is  not  affected  by  the
 details  of the  local geometry.
\vskip0.3cm

\subsection{Construction of the self-adjoint extension }

Extend $A_0$  from  $D_0$ onto $D (A^+_0)= D_0 + N_i + N_-i$ as an
``adjoint operator''  $A^+_0$  by setting  $\left(A^+_0 \pm
iI\right)g = 0$ for $g \in N_{\mp i}$. This  operator not
selfadjoint, and it is not even symmetric, so that  the boundary
form
\begin{equation}
\label{bform} \langle A_0^+  u,\, v\rangle - \langle  u,\, A_0^+
v\rangle = {\mathcal{J}}(u,v)
\end{equation}
does not vanish, generally, for  $ u,v \in D (A^+_0)$. One can
rewrite (\ref{bform}) in  more  convenient form with using  new
symplectic coordinates with respect of a new  basis in $N$
\[
W^+_s =  \frac{1}{2}\left[ g_s +  \frac{A+iI}{A-iI} g_s \right] =
\frac{A}{A -iI} g_s
\]
\[
W^-_s =  \frac{1}{2i}\left[ s_s - \frac{A+iI}{A-iI} g_s \right]=
-\frac{I}{A -iI} g_s
\]
Since $A_0^+ g_s  + i g_s = 0,\,\, [A_0^+  - iI] \frac{A+iI}{A-iI}
g_s = 0$ we  have,
\begin{equation}
\label{AW} A_0^+ W^+_s = W^-_s, \,\, A_0^+ W^-_s = - W^+_s.
\end{equation}
Following  \cite{Extensions}  we  will use  the  representation of
elements from the domain of the adjoint operator, by the expansion
on
 the new basis:
\[
 u = u_0 + \sum_{s} \xi^s_+ W^+_s  + \xi^s_- W^-_s = u_0 +
\frac{A}{A -iI} \sum_{s} \xi^s_+ g_s  - \frac{I}{A -iI} \sum_{s}
\xi^s_- g_s: =
\]
\begin{equation}
\label{symplectic} = u_0 + \frac{A}{A -iI} \vec{\xi}_{+} -
\frac{I}{A -iI} \vec{\xi}_{-}.
\end{equation}
Note that  due to (\ref{AW})
\[
A^+ \,\, \frac{A}{A - iI}\,\vec{\xi}_{+} = - \frac{I}{A - iI}
\vec{\xi}_{+},\,\,A^+  \, \frac{-I}{A - iI} \,\vec{\xi}_{-} = -
\frac{A}{A - iI} \vec{\xi}_{-}.
\]
Note that  the boundary  form
\[
\langle A_0^+ u, v \rangle - \langle u,A_0^+ v \rangle : =
{\mathcal{J}}(u,v)
\]
of   elements   $u,v$,
\begin{equation}
\label{uxipm} u = u_0 +  \frac{A}{A - iI}\vec{\xi}^u_{+} -
\frac{I}{A - iI}\vec{\xi}^u_{-}: = u_0 +  n^u,\, u_0 \in D(A_0) \,
n^u \in N,
\end{equation}
\[
v = v_0 +  \frac{A}{A - iI}\vec{\xi}^v_{+} - \frac{I}{A -
iI}\vec{\xi}^v_{-}: = v_0 +  n^v,\, u_0 \in D(A_0) \, n^v \in N
\]
depends  only  on  components $n^u, n^v$  of  $u,v$ in the defect
$N$. Then the boundary form is represented as:
\begin{equation}
\label{bound_form} \langle A_0^+ u, v \rangle - \langle u,A_0^+ v
\rangle : = {\mathcal{J}}(u,v) = \langle \vec{\xi}^u_+,
\vec{\xi}^v_-\rangle - \langle \vec{\xi}^u_-, \vec{\xi}^v_+\rangle
\end{equation}
with  Euclidean dot-product for  vectors  $\vec{\xi}_{\pm} \in N_i$.
Note  that the  representation of the  boundary form  in terms  of
abstract  boundary  values $\vec{\xi}_{\pm}$  contains  only
integral characteristics  of  the  elements  from the  domain of
 the  operators considered, and hence it is  stable with respect of minor
local perturbations  of  geometry of the plates. This enables us to
substitute, for  practical calculations, the  real irregular
boundaries of the plates  by  the  smoothed  boundaries, obtained
via elimination of minor geometrical details, compared with the
length of  standing waves,  circa 5000 km, of SGO with periods in
the essential  gange  20 min - 1 hour.

The boundary form vanishes on the Lagrangian plane defined in
$D(A^+_0)$ defined by the ``boundary condition'' with an Hermitian
operator  $M : N_i \to N_i$ :
\begin{equation}
\label{bcondM} \vec{\xi}_+ = M \vec{\xi}_-.
\end{equation}
This  boundary condition defines  a  self-adjoint operator  $A_M$ as
a  restriction of  $A_0^+$  onto the  Lagrangian plane ${\mathcal
{T}}_M \in D(A_0^+)$ defined  by the  boundary condition
(\ref{bcondM}). The resolvent of  $A_M$ defined by the boundary
conditions is  represented, at regular points of $A_M$, by the Krein
formula, see \cite{Akh_Glazman66,Extensions}:
\begin{equation}
\label{E:Krein1} \left( A_M - \lambda I \right)^{-1} = \frac{I}{ A -
\lambda I} - \frac{A+iI}{A -\lambda I} P  M \frac{I}{ I + P \frac{I+
\lambda A}{A - \lambda I} P M} P  \frac{A-iI}{A -\lambda I},
\end{equation}
where $P$ is  an orthogonal projection onto $N_i$.
\vskip0.3cm
\subsection{Compensation of singularities in Krein formula\\ and
calculation of the  perturbed  spectral  data}
 Singularities  of the resolvent $\left(A_M - \lambda I
\right)^{-1}$ coincide with the spectrum of $A_M$. But  both terms
in the right side of (\ref{E:Krein1}) also have  singularities on
the  spectrum of  the non-perturbed  operator $A$. {\it The
singularities of the first and second term  the eigenvalues of $A$
compensate each other}. We are able to  derive this statement via
straightforward calculation, in classical  Krein-Birman-Schwinger
formula  and in the corresponding construction for quantum networks,
see \cite{MP2007, APY_INI_07}. In the course of the calculation  of
the  compensation  of singularities  we can recover both the
eigenvalues of the perturbed operator $A_M$ and the corresponding
eigenfunctions, see \cite{MP2007}. Note that a similar statement, as
a lemma on compensation of singularities of the corresponding
Weyl-Titchmarsh function, was discovered in \cite{Ring} for 1-d
solvable model of the quantum network in form of a quantum graph.
Later, in \cite{MP01} and in \cite{MP02}, similar statements were
proven  for Dirichlet-to-Neumann maps of  quantum networks. We
formulate  here this statement for the resolvent of the selfadjoint
extension based on ideas proposed  in \cite{MPP04}.

We  will observe the  effect of compensation of singularities on a
certain spectral interval  $\Delta_0 = [\lambda_0-\delta,\,
\lambda_0 + \delta]$, centered at  the {\it resonance eigenvalue}
$\lambda_0$ of the non-perturbed plate, assuming that the
perturbation defined by the matrix  $M$ is relatively small, in a
certain sense, se below.

Assuming that there is a single  eigenvalue  $\lambda_0$ of $A$  on
the interval $\Delta_0$,  with the  eigenfunction $\varphi_0$, we
use  the  following representations, separating  the polar terms
from smooth operator functions $ K_i, K_{-1},K$  on $\Delta_0$
 \[
\frac{A + iI}{A - \lambda I} = (\lambda_0 + i)
 \frac{\varphi_0\rangle \langle \varphi_0}{\lambda_0 - \lambda} + K_{-i},
 \]
 \[
\frac{A - iI}{A - \lambda I} = (\lambda_0 - i)
 \frac{\varphi_0\rangle \langle \varphi_0}{\lambda_0 - \lambda} + K_{i},
 \]
\begin{equation}
\label{small} P \frac{I + \lambda A}{A - \lambda I}P = (1 +
\lambda_0^2) \frac{P \varphi_0\rangle \,\langle P \varphi_0
}{\lambda_0 - \lambda} +  K (\lambda ),
\end{equation}
with a smooth  matrix-function  $K
 = K_0 +  o(|\lambda -
\lambda_0|)$, with  $K_0 = K (\lambda_0)$ and  $\parallel o(|\lambda
- \lambda_0|)\parallel \leq C_{0} \delta$.

\begin{definition}\label{smallM} { We say that
 the matrix $M$ is  relatively small, if
$ [I +  K(\lambda)M ]^{-1} $  exists and is bounded on  $\Delta_0$.}
\end{definition}

This  condition is obviously fulfilled if
\begin{equation}
\label{smallMC} \parallel [I +  K(\lambda_0) M]^{-1}\parallel\,
C_{0} \delta << 1.
\end{equation}
To calculate the second term in the  right side of the Krein formula
(\ref{E:Krein1})  we  have to compute the inverse of the
denominator, that is to solve the  equation
\begin{equation}
\label{E:polar} \left[  I + P \frac{I+ \lambda A}{A - \lambda I}
M\right] u = g
\end{equation}
Though the  standard  analytic perturbation technique is  still not
applicable  to this  equation under the above  conditions
\ref{smallM} or (\ref{smallMC}), we  are able to  construct the
inverse  based on  finite  dimensionality (one-dimensionality) of
the  polar term.
\begin{equation}
\label{E:polar1} u = [I+ KM]^{-1}g - (1 + \lambda_0^2)\frac{[I+
KM]^{-1}P \varphi_0\rangle\langle P \varphi_0 M[I+
KM]^{-1}g\rangle}{\lambda_0 -\lambda + \langle P \varphi_0 M[I+
KM]^{-1}\varphi_0\rangle}.
\end{equation}
Based on (\ref{E:polar1}) we are able, see  \cite{MP2007,APY_INI_07}
to observe the compensation of singularities in the above Krein
formula (\ref{E:Krein1}) and calculate  the polar term  of the
resolvent at the  single eigenvalue  of the operator  $A_M$ on the
interval $\Delta_0$: \vskip0.3cm \begin{theorem}{If the perturbation
is relatively small, as required in (\ref{smallM}), then there exist
a single eigenvalue $\lambda_M$ of the perturbed operator $A_M$ on
the interval $\Delta_0$ which is found as a  zero of the denominator
in (\ref{E:polar1})
\begin{equation}
\label{eigenvM} \lambda_0 -\lambda + (1+\lambda_0^2)\langle
P\varphi_0 M[I+ KM]^{-1}\varphi_0\rangle:= {\bf d}_M(\lambda),\,
{\bf d}_M(\lambda_M) = 0
\end{equation}
and the corresponding eigenfunction
\begin{equation}
\label{eigenvP} \varphi_M = \varphi_0 - (\lambda_0 - i )K_{-i}M[I +
KM]^{-1}P \varphi_0,
\end{equation}
computed at the  zero  $\lambda_M$. The  polar term  of the
resolvent of the perturbed  operator at the  eigenvalue is
represented as:
\[
\frac{\varphi_0 - (\lambda_0 - i )K_{-i}M[I + KM]^{-1}P
\varphi_0\rangle \langle \varphi_0 - (\lambda_0 - i )K_{-i}M[I +
KM]^{-1}P \varphi_0}{{\bf d}_M(\lambda)}
\]
}
\end{theorem}
{\it Proof} \,\,\, of this  statement can be  obtained  similarly to
the  corresponding  statement in \cite{MP2007}, where the case of
several eigenvalues  of the  non-perturbed  operator on an essential
spectral interval was  discussed. We  consider here the simplest
case, when only one  eigenvalue of the  unperturbed  plate is
present on  an  essential spectral interval. If the perturbation is
small as required  by the condition (\ref{smallMC}), then the
approximate eigenvalue and the corresponding approximate
eigenfunction of $A_M$ can be obtained via replacement $K, K_{\pm
i}$ in (\ref{eigenvM},\ref{eigenvP})  by $K(\lambda_0), K_{\pm
i}(\lambda_0)$:
\begin{equation}
\label{data} \lambda_M \approx \lambda_0 + (1+\lambda_0^2)\langle
P\varphi_0 M[I+ K(\lambda_0)
 M]^{-1}K(\lambda_0) \varphi_0\rangle,
\]
\[
\varphi_M \approx \varphi_0 - (\lambda_0 - i )K_{-i} (\lambda_0)M[I
+ K(\lambda_0)M]^{-1}P \varphi_0.
\end{equation}

\noindent {\bf Remark} Analysis of the multi-point boundary
condition which corresponds to  several stresses  applied at the
points $a_1,a_2,\dots a_m$ on the  boundary of the  plate $\Omega$
differs from the above analysis of the single-point  case, only  in
the first step. In the case of a  multi-point  stress we  have to
construct of elements $\left\{ g_0^r(x,a_r,i),\,g_1^r(x,a_r,i),\,
g_2^r(x,a_r,i) \right\}_{r=1}^{m}$ a basis in the larger  deficiency
subspace $N_i$,\, dim $N_i = 3m$. Due to the presence of
singularities of different types at  different points, the
deficiency elements are linearly independent. \vskip0.3cm
\section{ Concluding remarks on the  fitting\\ of the  model }
The  pair  of data (\ref{data}) may be used in two different ways:
either for calculation of the shift  of the  frequency  of  SGO and
the  corresponding  perturbation of the eigenfunction, under the
point-wise  stress  characterized by  the matrix $M$, or, vice
versa, for  recovering of the data on  the localization, the  type
and the  intensity of the  stress from instrumental observations.

Indeed, if  the geological structure of the tectonic plates at the
active zones, encoded in  matrices $M_s$ attached to the zones, are
known, then, theoretically, we are able to calculate the
eigenfunctions  and  the eigenfrequencies of the plates, taking into
account the stress caused by  collisions. We are also able,
theoretically, to construct the deficiency elements for all active
zones. Then  the self-adjoint extension of the bi-harmonic operator
on the plate, with the point-wise boundary stress, can be
constructed, with  the corresponding matrices $M_s$. The obtained
theoretical results can be compared with  the results of  the
instrumental measurements. This permits to recover the matrices
$M_s$, which characterize the stressed  points $a_1,\, a_2,\,
\dots$.

Assume that the structure of the plates at the collision point $a$
remains unchanged, but the tension is growing linearly with time
$\tau$ as: $M (\tau) = m \tau$, with a matrix coefficient  $m: N \to
N$. Then the formulae (\ref{eigenvP},\ref{eigenvM}), for small
$\tau$, define the derivatives  of $\lambda_M,\varphi_M$ with
respect to $\tau$ at the moment $\tau = 0$:
\[
\frac{\partial \varphi_{M}}{\partial \tau}\bigg|_{\tau = 0} =
-(\lambda_0 - i) K_{-i} m_0 P_+ \varphi_0 (x_s),\,\,\,
\frac{\partial \lambda_M}{\partial \tau} \bigg|_{\tau = 0} =  (1 +
\lambda^2_0) \langle  P_+ \varphi_0 , m\,\,  P_+ \varphi_0 \rangle.
\]
Comparing  this  result with ratios
\[
\frac{\varphi_{m\tau}(x_s) - \varphi_0}{\tau}(x_s) \bigg|_{\tau =
0},\,\, \frac{\lambda_{m\tau} - \lambda_0}{\tau} \bigg|_{\tau = 0},
\]
measured  experimentally for the  amplitude  and  frequency  of SGO,
we are able to find fit $m$, and calculate the increment of $M$ $
\delta M = \tau m$.

Practical experience  in analytic perturbations  shows, that  minor
perturbations affect  rather  the eigenvalues, than the  the shapes
of the  eigenfunctions  of the  spectral problem.  Based  on this
observation we  can estimate  the  speed  of  accumulation of
elastic energy $\mathcal{E}$,  under   the point-wise  boundary
stress depending  on the speed of the shift of the  eigenvalues
(eigenfrequencies)  of SGO  and initial distribution   of  the
elastic  energy on the modes  $\varphi^s_0$ defined  by the
corresponding  Fourier coefficients $\langle u, \varphi^s_0
\rangle$:
\[
\frac{d {\mathcal{E}}}{d \tau} \approx \sum_s \frac{d \lambda_M^s}{d
\tau} |\langle u, \varphi^s_0 \rangle |^2 = \sum_s
 (1 +
(\lambda^s)^2_0) \langle  P_+ \varphi^s_0 , m\,\,  P_+ \varphi^s_0
\rangle\,\,  |\langle u, \varphi^s_0 \rangle |^2,
\]
with  the  summation extended only on the  eigen-modes which
correspond to the  varying eigenvalues. If only one active zone
$a_s$ is  involved at a time, then only one matrix $M_s$  has to be
taken into account, so that  the risk of the powerful earthquake may
be estimated based on the magnitude of $\delta _s = \tau m_s $.

If there are several active zones at the  points $a_1, a_2, \dots
a_n $ on the boundary  of the plate, then the corresponding matrices
$M_1,M_2,\dots M_n$ can be fitted based on  observations of SGO  in
the remote zone during  preceding earthquakes which occurred at
$a_1, a_2, \dots a_n $. Variations of the  frequencies and the shape
of SGO may arise from the stress at any active zone, but usually
only one active zone is involved at a time. Once the matrices
$M_1,M_2,\dots $ are known, then comparison of the perturbation of
the frequencies  and the shapes of  seismo-gravitational  {\it
modes}, SGM (or, probably, seismo-gravitational pulsations, SGP) at
the given groups of points in the remote zone with results of
previous measurements at these points, would  allow  us  to identify
the active zone where the stress is applied. We presume that the
above model gives a chance to introduce a useful system into the
scope of the experimental data on the seismo-gravitational
oscillations in remote zone and use them for estimating of risk and
localization of the powerful earthquakes. This  opens  an
alternative  to the statistical methods, see \cite{GNS05},  of
estimation  of risk of powerful earthquakes.

Fitting of the proposed model in reality requires both extended
computing and a major experimental data base. Because the
wavelengths of the  standing waves,  circa 5000 km, on the essential
range of periods,  20 min- 1 hour,dominate the size of the  active
zones, one may assume, that the straightforward computing with
averaged and smoothed data for Young's modulus and geometric
characteristics of the plates will enables us to obtain  a realistic
approximation of the deficiency elements, with singularities at the
active zones, and to construct the perturbed eigenfunctions of the
plates,  which correspond to SGO.

More accurate theory requires  taking into account  realistic
boundary conditions, and the  exchange of energy with the liquid
underlay and neighboring plates. In particular, arising  new modes
in the  spectrum of SGO  of  the plate, transferred, due to the
tight contact in active zone, from the neighboring plate may be
considered as another possible  precursor of a  powerful earthquake.
The choice of realistic boundary conditions has to be done based on
experimental data interpreted within an appropriate extension of the
scheme proposed above,  with Lame and hydro-dynamical equations
involved. We  postpone discussion of these interesting questions to
oncoming publications.

\section{Acknowledgement} The initial version of the text  was
presented at the I.G.Petrovskii conference Moscow, May 21-26, 2007,
see \cite{P07}, and issued as a preprint  of the  Semester of
Quantum Graphs at the International Newton institute, Cambridge,
January - June 2007, see \cite{PP07}. The authors acknowledge
support from the organizers and participants of these  meetings. The
authors are grateful to  Doctor  Colin Fox  for  an inspiring
discussion  of the Saint-Venant principle. L. Petrova acknowledges
support from the RFFI grant 01-05-64753, B. Pavlov  acknowledges
support from the grant of the Russian Academy of Sciences, RFFI
97-01-01149.

\end{document}